\documentstyle[aps,preprint,epsfig,amssymb,floats]{revtex}

\begin{document}
\draft
\preprint{UAB--FT--447}
\title{\Large {\bf Long Range Forces From Pseudoscalar Exchange}}

\vspace{2in}
\author{F. Ferrer and J. A. Grifols}

\vspace{.5in}

\address{Grup de F{\'\i}sica Te\`orica and Institut de F{\'\i}sica d'Altes Energies,\\
Universitat Aut\`onoma de Barcelona, E--08193 Bellaterra, Barcelona, Spain}
\date{May 1998}
\maketitle

\vspace{1in}

\begin{abstract}
Using dispersion theoretic techniques, we  consider coherent long range
forces arising from double pseudoscalar exchange among fermions. We find
that Yukawa type coupling leads to $1/r^3$ spin independent  attractive
potentials whereas derivative coupling renders $1/r^5$ spin independent
repulsive potentials.
\end{abstract}

\vspace{1in}

\section{Introduction}
\label{sec:intro}

 Many extensions of the Standard model predict the existence of light
scalar particles. The axion may be the most debated one but there are
also approximate Nambu-Goldstone fields associated with family
symmetries, or moduli fields, or dilatons, or superpartners of the
gravitino~\cite{gelmini,dvali,wil,rafa}. Exchange of such particles by ordinary matter will induce
forces that extend over the Compton wavelength of the particle~\cite{dvali,moody,fishbach}.
However, the effect will be felt by bulk matter only if the potential is
spin-independent so that forces can add up coherently over macroscopic
distances. Now, a pseudoscalar particle, such as the axion, is
coupled to fermions via a $\gamma_{5}$ which, in the
nonrelativistic limit, flips the spin. Therefore, single pseudoscalar
exchange leads to spin-dependent forces that do not extend over
macroscopic unpolarized bodies~\cite{wil,moody}. A double exchange of pseudoscalars
on the other hand can coherently sum over a macroscopic sample
of matter because it can leave the spin unflipped. The explicit form of
these forces has been derived and their phenomenological consequences
explored in previous work in the context of nonrelativistic "old
fashioned perturbation theory"~\cite{grifols}. Here we reopen the question of
pseudoscalar mediated forces in the light of the powerful dispersion
theoretical techniques devised by Feinberg and Sucher and collaborators
that make extensive use of full relativistic quantum field theory~\cite{fsarep,fsb}. 
 
In section~\ref{sec:metode} we give the necessary theoretical background which
is nothing but a brief resum\'e of the seminal work by Feinberg and Sucher.
Section~\ref{sec:yuk} is devoted to the Yukawa type interaction and
section~\ref{sec:der} deals with derivatively coupled scalars. We shall see
that the two interactions produce quite different potentials~\cite{ferrer}, a
fact that could not have been derived in a purely non-relativistic
framework~\cite{tortosa}. The paper ends with a brief
summary and conclusions contained in section~\ref{sec:conc}.

\section{Dispersive forces}
\label{sec:metode}
Following the general strategy devised by Feinberg and Sucher~\cite{fsarep,fsb}, we
define a potential in a given Quantum Field Theory by equating the
scattering amplitude for a two body process that follows from the usual
Feynman rules, with the transition amplitude associated to a Schr\"odinger
type equation solved {\it \`a la} Lippmann-Schwinger. Let us be explicit and
consider elastic scattering of particles A and B with four momenta
$p_{a}$ and $p_{b}$ in the initial state and $p_{a}'$ and $p_{b}'$ in
the final state. The Mandelstam variables are then:

\begin{equation}
s={\left( p_a+p_b \right)}^2\;\;\;\;\;t=Q^2\;\;\;\;\;u={\left( p_a-p_b
    ' \right)}^2
\end{equation}
with $Q=p_a-p_a'=-p_b+p_b '$.

In the C.M. we write the momenta as
\begin{equation}
\begin{array}{cc}
p_a=\left( E_a,{\bf p} \right) & p_b=\left( E_b,-{\bf p}
\right)\\
p_a'=\left( E_a,{\bf p}\,{}'
\right) & p_b'=\left( E_b,-{\bf p}\,{} ' \right).
\end{array}
\end{equation}

Now $s=W^2$ where $W=E_{a}+E_{b}$ and $t=-{\bf Q}^2$ with
$Q=(0,{\bf Q})$, whose physical region is 
\begin{equation}
s\geq s_0\;\;\;\;and\;\;\;\;-4 {\bf p}^2 \leq t \leq 0 \label{eq:rf}
\end{equation}
where 
\begin{equation}
s_0={\left( m_a+m_b \right)}^2\;\;\;\;\;{\bf p}^2=\left. \left[ s-{\left(
        m_a+m_b \right)}^2 \right] \left[ s-{\left(
        m_a-m_b \right)}^2 \right] \right/ 4 s.
\end{equation}

The transition from initial state $i$ to final state $f$ is described
in  Quantum Field Theory by the transition matrix element 
\begin{equation}
T_{fi}=N_f {\cal M}_{fi} N_i
\end{equation}
where $N_{f,i}$ are normalization factors of one particle states and ${\cal
  M}_{fi}$ is
the invariant Feynman amplitude.

The definition of our potential follows now from identifying this
transition amplitude with
\begin{equation}
T_{fi}=\left\langle {\bf p}',-{\bf p}' \right|V+V\,{\left( W-h_0-V+i\epsilon
  \right)}^{-1}\,V \left|{\bf p},-{\bf p} \right\rangle   \:\frac{m_a
  m_b}{E_a E_b}
\end{equation}
where $h_{0}$ is the sum of the free Dirac Hamiltonians for particles A
and B. Here all quantities are referred to the C.M.

The Feynman amplitude ${\cal M}$ is understood as a series expansion in (even)
powers of the coupling constant associated to single, double, \ldots
particle exchange in the t-channel. We assume that the potential V
also admits a series expansion 
\begin{equation}
V=V^{(2)}+V^{(4)}+...\;\;.
\end{equation}

So we determine V order by order in perturbation theory through
\begin{eqnarray}
\left\langle {\bf p}',-{\bf p}' \right| V^{(2)} \left|{\bf p},-{\bf p}
\right\rangle  & = & {\cal M}^{{(2)}}_{fi}  \\
\left\langle {\bf p}',-{\bf p}' \right| V^{(4)} \left|{\bf p},-{\bf p}
\right\rangle & = & {\cal M}^{(4)}_{fi}-\left\langle {\bf p}',-{\bf p}'
\right| V^{(2)}\, \left(
     W-h_0+i\epsilon \right)^{-1} V^{(2)}\left|{\bf p},-{\bf p}
\right\rangle .
\end{eqnarray}

Here we should point out a technicality. The potential we are after must
have the form~\cite{min}
\begin{equation}
V=\Lambda_{++}\,U\, \Lambda_{++}
\label{eq:uv}
\end{equation}
where $\Lambda_{++}=\Lambda_{+;a}\Lambda_{+;b}$ is an operator that
projects on the positive energy states of $h_{0}$. Because
$\Lambda_{++}\mid{\bf p},-{\bf p}\rangle=\mid{\bf p},-{\bf p}\rangle$
and $\Lambda_{++}^2=\Lambda_{++}$ we can rewrite the previous equations
that determine the potential as,
\begin{eqnarray}
\left\langle {\bf p}',-{\bf p}' \right| U^{(2)} \left|{\bf p},-{\bf p}
\right\rangle &=&{\cal M}^{(2)}_{fi} \label{eq:u2} \\
\left\langle {\bf p}',-{\bf p}' \right| U^{(4)}\left|{\bf p},-{\bf p}
\right\rangle &=&{\cal M}^{(4)}_{fi}-\left\langle {\bf p}',-{\bf p}' \right| U^{(2)}\,\left( W-h_0+i\epsilon
  \right)^{-1}\,\Lambda_{++}\, U^{(2)}\left|{\bf p},-{\bf p}
\right\rangle  \label{eq:u4}\\
\vdots \nonumber
\end{eqnarray}

 In principle the above equations permit an iterative determination of
the potential to the desired order. However we are not done yet because
we would like to have our potential in position space and what we have
is the operator $U$ in the momentum representation. Therefore, we should
Fourier transform our results back to configuration space, i.e. we wish
to find $U^{(n)}(r)$ such that
\begin{equation}
\left\langle {\bf p}',-{\bf p}' \right| U^{(n)}\left|{\bf p},-{\bf p}
\right\rangle =\int  d{\bf r}\; e^{i{\bf Q}\cdot {\bf r}} \:U^{(n)}(r) .
\end{equation}

Inversion of the above equation, however, requires knowing the function
${\cal M}$ for all values of three momentum. But we only know the scattering
amplitude on shell, i.e. for ${\bf p}^{2}={\bf p}'^{2}$. We can use the
fact that ${\cal M}(s,t)$ is an analytic function of t and so analytically extend
its domain beyond the physical region, i.e. for all values of
${\bf Q}^2=t$.

Suppose ${\cal M}(s,t)$ is analytic everywhere except for branch cuts on the real
axis and furthermore, it vanishes for large $\mid t \mid $. Then, using
Cauchy's theorem, we can write
\begin{eqnarray}
{\cal M}^{(R)}&=&\frac{1}{\pi} \int^{\infty}_{t_0}{d t'\; \frac{\rho^{(R)}
    (s,t')}{t'-t}}\label{eq:espr}\\
{\cal M}^{(L)}&=&\frac{1}{\pi} \int_{-\infty}^{\bar{t}_0}{d t' \; \frac{\rho^{(L)}
    (s,t')}{t'-t}}\label{eq:espl}
\end{eqnarray}
where $\rho (s,t)=\frac{\left[ {\cal M} \right]_t}{2 i}$ is the spectral
density and $\left[ {\cal M}
\right]_t$ is the discontinuity of ${\cal M}$
across the cut. Only the piece of the amplitude arising from the right
hand cut will be of interest to us for only this piece leads to a long
range potential~\cite{fsd88}. Assuming that the basic relations eqs~(\ref{eq:u2})~and~(\ref{eq:u4}) hold also in the extended domain, we can Fourier invert
them as follows\footnote{Notice that our generalized potential
will depend on the parameter $s$.},
\begin{equation}
U^{(n)} (r;s)=\frac{1}{\left( 2 \pi
  \right)^3}\int{\!\!d{\bf Q}\;\:e^{-{\bf Q}\cdot{\bf r}} \:{\cal M}^{(n)} \left(
    s,-{\bf Q}^2 \right)}. \label{eq:fou}
\end{equation}

We use now the spectral representation given before to obtain
\begin{equation}
U^{(n)} (r;s)=\frac{1}{4 \pi^2 \: r} \;\int^{\infty}_{t_0}{\!\!dt\; \rho^{(n)}
  (s,t)\;e^{-\sqrt{t}\, r}}
\label{eq:pott}
\end{equation}
where, to reach this final form, we conveniently changed the order in
which integrals were done. In short, obtaining long range potentials
amounts to calculating t-channel discontinuities in Feynman diagrams
and performing a Laplace transform. We shall see how things work out
in detail as we do our specific calculations in the next two sections.

\section{The Yukawa coupling}
\label{sec:yuk} 
 Our starting point is the Lagrangian density,
\begin{equation}
{\cal L}^{Y}_{int}=-i g\, \bar{\Psi} (x) \gamma^5 \Psi (x) \,\Phi (x)
\label{eq:lps}
\end{equation}
where $\psi$ is a fermion field and $\phi$ is the pseudoscalar field
which we take to be massless\footnote{For scalars with mass, the long
distance potentials are damped with Yukawa exponentials. In this case,
our results are valid for distances on the order or smaller than the
Compton wavelength of the exchanged particles.}. The potential
associated to single particle exchange is easily obtained from the
discontinuity associated to the diagram in Fig. 1. The spectral density
function
$\rho^{(2)}(s,t)$ is in this case
\begin{equation}
\rho^{(2)} (s,t)=\pi g^2 \;\bar{u} \left( p_a'{} \right) \gamma^5 u\left( p_a
\right) \: \bar{u} \left( p_b'{} \right) \gamma^5 u\left( p_b \right)\:
\delta (t) .
\end{equation}

After Laplace transforming we get the relativistic potential operator
\begin{equation}
U^{(2)}=\frac{g^2}{4 \pi \; r}\; \gamma^0_a \gamma^5_a\:\gamma^0_b \gamma^5_b
\label{eq:u2ps}
\end{equation}
where subindices make explicit that Dirac matrices act either on spinor
A or spinor B. 
 
 The nonrelativistic limit of equation above leads to the well known
spin-dependent potential

\begin{equation}
V^{(2)}_{nr}=\frac{g^2}{4 \pi \;r \, \left( 2 m_a \right) \left( 2 m_b
  \right)}\;\;
{\mbox{\boldmath $\sigma$}}_a \!\!\cdot\!\! {\mbox{\boldmath $\nabla$}}\, \otimes \,{\mbox{\boldmath $\sigma$}}_b \!\!\cdot\!
\!{\mbox{\boldmath $\nabla$}} 
\label{eq:v2ps}
\end{equation}
with $m_{a}$ and $m_{b}$, the masses of particles A and B respectively.
What we are really interested in is $U^{(4)}$, i.e. the potential due to
two-particle exchange. To this end we need the discontinuities of
diagrams in Fig. 2 and the discontinuity of the subtraction term in
eq~(\ref{eq:u4}) (iteration of the lowest order
potential $U^{(2)}$).

The Feynman amplitude associated to Fig. 2 can be written
\begin{eqnarray}
{\cal M}^{(4)}  = \frac{i}{2! (2 \pi)^4}\; \int{\!\!d^4 k \,d^4
  k'{}\;\delta^{(4)}(Q-k-k'{})  \:
  \frac{1}{k^2+i\epsilon}\frac{1}{{k'{}}^2+i\epsilon}} \nonumber \\
{\cal M}^C_a \left( -k,k'{};P_a
  \right) {\cal M}^C_b \left( k,-k'{};P_b \right) \label{eq:m4c}
\end{eqnarray}
in terms of the Compton amplitude, depicted in Fig. 3,
\begin{equation}
{\cal M}^C \left( k,k'{};P \right)=g^2 \; \bar{u} (p '{}) \left[
  \frac{\rlap / k}{2\: p\cdot k} +\frac{\rlap / k'}{2\: p\cdot k'{}}
\right] u(p) .\label{eq:comps}
\end{equation}

Making use of the Dirac equation and trading the propagators for their
discontinuities, i.e.
\begin{equation}
\frac{1}{k^2+i\epsilon}\longrightarrow -2 \pi i\, \delta \left( k^2 \right)
\Theta \left( k^0 \right) 
\end{equation}
we arrive at
\begin{equation}
\left[{\cal M}^{(4)}\right]_t=-\frac{i g^4}{8 \pi^2}\;\int{\!d\Phi\;\bar{u}_a'{} \left[
  \frac{p_a\cdot(k'{}-k)\: \rlap / k}{2\: p_a \cdot
    k'{} \: p_a\cdot k} \right] u_a\;\bar{u}_b'{} \left[\frac{p_b\cdot (k'{}-k)
   \: \rlap / k}{2\: p_b \cdot k'{} \: p_b\cdot k}\right]
u_b}
\end{equation}
with the two particle phase space explicitly given by
\begin{equation}
d\Phi=\delta \left(Q-k-k'{} \right)\delta \left(k^2\right)\delta \left({k
  '{}}^2 \right) \Theta  \left(k^0\right) \Theta  \left({k'{}}^0\right)\: d^4k\,
d^4k'{} .
\end{equation}

It is convenient to do the integrals in the C.M. of the pseudoscalars,
i.e. to go to the t-channel, and then use crossing symmetry to recover
the original amplitude. We follow here the notation in reference~\cite{fsd92} where
they deal with a related problem. First define momenta as,
\begin{equation}
\begin{array}{cc}
p_a=\left( \frac{\sqrt{t}}{2},{\bf p} \right) & p_{\bar{a}}=-p_a '{}=\left(
  \frac{\sqrt{t}}{2} ,-{\bf p} \right)\\
p_{\bar{b}}=-p_b=\left( \frac{\sqrt{t}}{2},-{\bf p}\,{} '{} \right)&  p_b
  '{}=\left( \frac{\sqrt{t}}{2} ,{\bf p}\,{} '{} \right)\\ 
k=\left( \frac{\sqrt{t}}{2},{\bf k} \right)& k '{}=\left(
  \frac{\sqrt{t}}{2} ,-{\bf k} \right)\label{eq:cin}
\end{array}
\end{equation}
introduce next the unit imaginary vectors
\begin{eqnarray}
{\bf p}&=&i\: \xi_a m_a\: \hat{{\bf p}}\\
{\bf p}\,{}'{}&=&i\: \xi_b m_b\: \hat{{\bf p}}'{}
\end{eqnarray}
with $\xi_{a,b}\equiv\sqrt{1-\frac{t}{4\: m^2_{a,b}}}$  and  $\hat{{\bf
    p}},\,\hat{{\bf p}} '{}$ are unitary complex vectors verifying $\hat{{\bf p}}\cdot\hat{{\bf p}}=-1$ so that
all particles are on-shell.

Now the discontinuity can be put in the form
\begin{equation}
\left[{\cal M}^{(4)}\right]_t=\frac{i g^4}{4 \pi \: b t}\;\int{\frac{d\Omega}{4 \pi}\; \frac{x_a
    x_b}{d_a d_b}\;\; \bar{u}_a'{} \rlap / k u_a\: \bar{u}_b'{} \rlap / k u_b} 
\end{equation}
where, to simplify expressions, we use
\begin{eqnarray}
&b \equiv m_a \xi_a\: m_b \xi_b&\nonumber\\
&x_a\equiv\hat{{\bf p}} \cdot \hat{k}\;\;\;\;\;x_{b} \equiv \hat{{\bf p}}
'{}  \cdot \hat{k}&\nonumber\\
&d_{a,b}\equiv\tau^2_{a,b}+x^2_{a,b}& \nonumber \\
&\tau_{a,b}\equiv\frac{\sqrt{t}}{2\: \xi_{a,b}\: m_{a,b}}&.\label{eq:defdab}
\end{eqnarray}

 The integration to be carried out is an angular average. We use the
shorthand: $\int {\!d\Omega \over {4\pi}}f\equiv \langle f\rangle$. Hence, the
discontinuity takes the form
\begin{equation}
\left[{\cal M}^{(4)}\right]_t=\frac{i g^4}{4 \pi \:b t}\; \bar{u}_a'{} \gamma_{\mu} u_a\; \bar{u}_b'{}
\gamma_{\nu} u_b \:{\cal T}^{\mu \nu}\label{eq:dism4ps}
\end{equation}
with ${\cal T}^{\mu\nu}\equiv \left\langle \frac{x_a x_b}{d_a d_b}\:k^{\mu}
  k^{\nu} \right\rangle$.

Lorentz covariance dictates the following decomposition
\begin{eqnarray} 
{\cal T}^{\mu\nu}&=&a_1\,P_a^{\mu} P_a^{\nu}+a_2 P_b^{\mu} P_b^{\nu}+a_3 \left( P_a^{\mu} P_b^{\nu} +
  P_b^{\mu} P_a^{\nu} \right)+a_4 g^{\mu\nu} \nonumber\\
& &\mbox{}+a_5 Q^{\mu} Q^{\nu}+a_6 \left( Q^{\mu} P_a^{\nu}+P_a^{\mu} Q^{\nu} \right)+a_7 \left(
  Q^{\mu} P_b^{\nu}+P_b^{\mu} Q^{\nu} \right) \label{eq:lor}
\end{eqnarray}
in terms of the three independent momenta
\begin{equation}
P_a\equiv p_a+p_a'{} \;\;\;\;\;P_b\equiv p_b+p_b'{}\;\;\;\;\;Q \equiv
k+k '{}.
\end{equation}

The coefficients $a_i$ can be found to be combinations of scalar
integrals as shown in the appendix. Now, in the C.M. of the incident
particles, the relations 
\begin{eqnarray}
\bar{u}_a'{} \rlap / {P_a} u_a \; \bar{u}_b'{} \rlap / {P_a} u_b&=&4\: m_a \: \bar{u}_a'{} u_a \: \bar{u}_b'{} \left( W
  \gamma_0 - m_b \right) u_b \nonumber\\
\bar{u}_a'{} \rlap / {P_b} u_a \; \bar{u}_b'{} \rlap / {P_b} u_b&=&4\: m_b\:\bar{u}_a'{} \left( W
  \gamma_0 - m_a \right) u_a \: \bar{u}_b'{} u_b\nonumber\\
\bar{u}_a'{} \rlap / {P_a} u_a \; \bar{u}_b'{} \rlap / {P_b} u_b&=&4\: m_a m_b \: \bar{u}_a'{} u_a \: \bar{u}_b'{}
u_b\nonumber  \\
\bar{u}_a'{} \rlap / {P_b} u_a \; \bar{u}_b'{} \rlap / {P_a} u_b&=&4\: \bar{u}_a'{} \left( W
  \gamma_0 - m_a \right) u_a \: \bar{u}_b'{} \left( W
  \gamma_0 - m_b \right) u_b\nonumber\\
\bar{u}_a'{} \rlap / {Q} u_a &=&0\nonumber\\
\bar{u}_b'{} \rlap / {Q} u_b &=&0
\end{eqnarray}
are easily established with the help of the Dirac equation. This leads
directly to 
\begin{eqnarray}
\left[{\cal M}^{(4)}\right]_t = \frac{i g^4}{4 \pi \:b t} \;\bar{u}_a'{}
\bar{u}_b'{} \left[4 m_a m_b \left( 2 a_3-a_1-a_2 \right)+4 m_b W \gamma_a^0
  \left( a_2-a_3 \right) \right. \nonumber\\
+ \left. \mbox{}4 m_a W \gamma_b^0 \left( a_1-a_3 \right)+\gamma_a^0
  \gamma_b^0 \left( 4 W^2 a_3+a_4 \right)-{\mbox{\boldmath $\gamma$}}_a
  {\mbox{\boldmath $\gamma$}}_b\, a_4 \right] u_a u_b. 
\label{eq:disa}
\end{eqnarray}

This discontinuity is a complex function since the $a_i$ are complex and
hence adds an imaginary component to the spectral density which would
finally contribute an imaginary piece to the potential. Inspection of
equation~(\ref{eq:disa}) immediately tells us that the offending piece comes
from the imaginary parts of the $a_i$. But we should recall that we still
have to subtract the contribution from the iterated lowest order
potential. It turns out that its imaginary part exactly cancels the
unwanted contribution coming from equation~(\ref{eq:disa}). Indeed we have
explicitly checked this to be the case. However, in order to make this
paper not too lengthy, we do not include the intermediate steps of the
calculation. We only report on the result, i.e.
\begin{eqnarray}
\Re\,\left[{\cal M}^{(4)}\right]_t & =&\frac{g^4}{8 p W} \;\bar{u}_a'{}
    \bar{u}_b'{} \left[\frac{t-4 p^2}{(4 p^2+t)^2} \left( E_a \gamma_a^0-m_a
    \right) \left( E_b \gamma_b^0-m_b \right)\right. \nonumber\\
& & - \left.\mbox{}\frac{p^2}{4 p^2+t} {\mbox{\boldmath $\gamma$}}_a \cdot
  {\mbox{\boldmath $\gamma$}}_b \right] u_a u_b .\label{eq:rm4ps}
\end{eqnarray}

The relevant contribution to the long range potential comes from the
real parts of the $a_i$, that is the imaginary part of~(\ref{eq:disa}), once
the contribution of the iterated potential has been subtracted. Let us
elaborate on the iteration amplitude,
\begin{equation}
{\cal M}_I=\left\langle {\bf p}',-{\bf p}' \right| U^{(2)}\,\left(
  W-h_0+i\epsilon \right)^{-1}\,\Lambda_{++}\, U^{(2)} \left|{\bf p},-{\bf p}
  \right\rangle .\label{eq:dmi}
\end{equation}

This formal expression can be recast in the explicit form\footnote{This
integral as it stands is infrared divergent. A fictitious mass regulator
is understood to be introduced in the scalar propagators which is set
to zero after the integrations are performed.} 
\begin{eqnarray}
{\cal M}_I &= &\frac{g^4}{8 \pi^2} \;\int{\! l^2 d l \,\int{ \!\frac{d \Omega}{4 \pi} \;u_a^{\dagger}{}'
    u_b^{\dagger}{}' \, \gamma_a^0 \left(E_a' \gamma_a^0-{\mbox{\boldmath $\gamma$}}_a \cdot {\bf l}
      - m_a \right)}}\nonumber\\
& & \gamma_b^0 \left(E_b' \gamma_b^0+{\mbox{\boldmath $\gamma$}}_b \cdot
      {\bf l} - m_b \right)\, u_a u_b \; {\cal C} (p,l) \: \frac{1}{{\bf q}\,{}'{}^2} \frac{1}{{\bf q}^{\,2}} 
\end{eqnarray}
with ${\bf q}\equiv {\bf p}-{\bf l}$ and ${\bf q}' \equiv {\bf p}'-{\bf l}$
  and ${\cal C} (p,l)\equiv \frac{1}{E_a ' E_b '
  (W-W'+i\epsilon)}$.

The integration over momentum ${\bf l}$ reflects the fact that we have
inserted a complete set of plane wave intermediate states in~(\ref{eq:dmi}).
We have also used
\begin{equation}
\Lambda_{+;a}({\bf l})=\frac{E'_{a}+\gamma_{a}^{0}{\mbox{\boldmath $\gamma$}}_a\cdot{\bf l}+\gamma_{a}^{0}m_{a}}{2E'_{a}}
\end{equation}
where $E'_{a}=\sqrt{m_{a}^{2}+l^{2}}$.

The iteration amplitude can be conveniently put as follows
\begin{eqnarray}
{\cal M}_I & = &\frac{g^4}{8 \pi^2}\; \int{\! l^2 d l \;\bar{u}_a'{} \bar{u}_b'{}\, \left\{ \left(E_a
      ' \gamma_a^0-m_a \right) \left(E_b ' \gamma_b^0-m_b \right) \,
    {\cal L}-{\mbox{\boldmath $\gamma$}}_a \cdot {\mbox{\boldmath ${\cal V}$}} \left(E_b ' \gamma_b^0-m_b
    \right) \right. }\nonumber\\
&  & +\mbox{}\left. \left(E_a ' \gamma_a^0-m_a \right) {\mbox{\boldmath $\gamma$}}_b \cdot
    {\mbox{\boldmath ${\cal V}$}} - \gamma_a^i \gamma_b^j \Upsilon_{i j} \right\} u_a u_b \;{\cal
    C} (p,l) 
\end{eqnarray}
where
\begin{eqnarray}
{\cal L}&\equiv& \int{ \frac{d\Omega}{4 \pi}\;\frac{1}{{\bf q
        }\,{}'{}^2} \frac{1}{{\bf q}^{\,2}}} \label{eq:dl}\\
{\mbox{\boldmath ${\cal V}$}}& \equiv &\int{ \frac{d\Omega}{4 \pi} \;{\bf l} \;\frac{1}{{\bf q
       }\,{}'{}^2} \frac{1}{{\bf q}^{\,2}}}\label{eq:dv}\\
\Upsilon_{i j}&\equiv& \int{ \frac{d\Omega}{4 \pi}\; l_i l_j\;\frac{1}{{\bf q
        }\,{}'{}^2} \frac{1}{{\bf q}^{\,2}}}\label{eq:dup}
\end{eqnarray}
and they can be found in the appendix.

A little bit of Dirac algebra and the results in the appendix  allow us to write the
discontinuity of ${\cal M}_I$ as 
\begin{eqnarray}
\left[{\cal M}_I\right]_t&=&\frac{g^4}{8 \pi^2} \;\int{ \!l^2 d l\;
  \bar{u}_a'{} \bar{u}_b'{} \left\{ \left(E_a  ' \gamma_a^0-m_a \right)
  \left(E_b ' \gamma_b^0-m_b \right)  \right. }\nonumber\\
&& -\mbox{} 2 \:\frac{p^2+l^2}{4 p^2+t} \left[\left(E_a \gamma_a^0-m_a \right)
  \left(E_b ' \gamma_b^0-m_b \right)+\left(E_a' \gamma_a^0-m_a \right)
  \left(E_b \gamma_b^0-m_b \right) \right]\nonumber\\
& & +\frac{4}{4 p^2+t} \left( 2 \frac{(p^2+l^2)^2}{4 p^2+t}-l^2 \right) \left(E_a
   \gamma_a^0-m_a \right) \left(E_b  \gamma_b^0-m_b \right)\nonumber\\
&&+\left.{\mbox{\boldmath $\gamma$}}_a  \cdot {\mbox{\boldmath $\gamma$}}_b
  \left(  \frac{(p^2+l^2)^2}{4 p^2+t}-l^2 \right) \right\} \left[{\cal
  L}\right]_t u_a u_b  \; {\cal C} (p,l).
\label{eq:cmi}
\end{eqnarray}

The explicit form for $[{\cal L}]_{t}$ is given in the appendix. Note that
in~(\ref{eq:cmi}) we have 
\begin{equation}
\frac{1}{W-W'+i \epsilon}=\wp \left( \frac{1}{W-W '} \right)-i \pi\:
\delta (W-W ') .\label{eq:pprin}
\end{equation}

The Dirac delta piece gives a contribution that, as already  advertised, will exactly cancel
the real part of the fourth order discontinuity function~(\ref{eq:rm4ps}). The principal part
integral can be cast in the form
\begin{equation}
i \Im \left[{\cal M}_I\right]_t=\frac{i g^4}{16 \pi
  \sqrt{t} (4 p^2+t)^2}\; \wp \int_{-1}^1 {\frac{d x}{\sqrt{1-x^2}}\: {\cal C} (p,
  l) \;\bar{u}_a'{} \bar{u}_b'{} \, {\cal N}} u_a u_b \label{eq:midisn}
\end{equation}
where we changed the integration variable via the relation
\begin{equation}
l^2=\frac{a'+b ' x}{2}\equiv \frac{1}{2} \left[ (2 p^2+t)+ \sqrt{t
    (4 p^2+t)} x \right] \label{eq:defx}
\end{equation}
and we used the shorthand
\begin{eqnarray}
{\cal N} &\equiv& (4 p^2+t)^2 \left(E_a ' \gamma_a^0-m_a \right) \left(E_b
  ' \gamma_b^0-m_b \right)-2 (4 p^2+t) (p^2+l^2)\nonumber \\
&&\left[\left(E_a
    \gamma_a^0-m_a \right) \left(E_b ' \gamma_b^0-m_b \right)+\left(E_a
    ' \gamma_a^0-m_a \right) \left(E_b \gamma_b^0-m_b \right) \right]\nonumber\\
&&+\mbox{}4
 \left( 2 (p^2+l^2)^2-l^2 (4 p^2+t) \right) \left(E_a \gamma_a^0-m_a
\right) \left(E_b \gamma_b^0-m_b \right)\nonumber\\
&&-\mbox{}{\mbox{\boldmath $\gamma$}}_a \cdot {\mbox{\boldmath $\gamma$}}_b
(4 p^2+t) \left( (4 p^2+t) l^2-(p^2+l^2)^2 \right).\label{eq:n}
\end{eqnarray}

It is convenient now to split the function ${\cal C}(p,l)$ as
\begin{eqnarray}
{\cal C} (p,l) &\equiv& {\cal C}_1 (p,l) + {\cal C}_2 (p,l) \nonumber\\
&\equiv& \frac{2}{W
  (p^2-l^2)} +\frac{1}{E_a ' E_b ' W} \left(
  \frac{p^2+l^2+m_a^2+m_b^2}{E_a E_b+E_a ' E_b '}+ \frac{W
    '}{W+W '} \right).\label{eq:sepc}
\end{eqnarray}

The integral above cannot be done exactly and we will expand the integral in a power series
in $t$ and $p^2$. This is a licit procedure because we will perform a Laplace transform that
heavily weighs the small t region of the spectral function when determining the long
range~(large r) potential and, we will eventually take the non relativistic limit of the
potential, i.e. for $p^2\sim 0$. Furthermore, each extra power of $t$ or $p^2$ implies a
correction to the potential with an extra power of $r^{-1}$. We see from equation~(\ref{eq:sepc})
that the calculation will involve doing integrals of the type
\begin{equation}
I_{c1} (n)\equiv\:\int_{-1}^1 { \frac{d x}{\sqrt{1- x^2}} \frac{l^n}{p^2-l^2}}
\label{eq:ic1}
\end{equation}
\begin{equation}
I_{c2} (n)\equiv\:\int_{-1}^1 { \frac{d x}{\sqrt{1- x^2}}\: l^n}.
\label{eq:ic2}
\end{equation}

The explicit results of the integrals needed in our calculation are also given in the appendix.
Armed with all this artillery we find for the ${\cal C}_1$ piece of the
discontinuity $[{\cal M}_I]_t$:
\begin{eqnarray}
i \Im \left[{\cal M}_I^{C_1}\right]_t&=&\frac{i g^4}{8 \sqrt{t} (4 p^2+t)^2
  (m_a+m_b)}\;\bar{u}_a'{} \bar{u}_b'{} \nonumber \\
&&\left\{ m_a m_b  \left[ 4 t- \frac{2 t p^2}{ m_a m_b}+\frac{p^4 t}{2
      m_a m_b} \left( \frac{1}{m_a^2}+\frac{1}{m_b^2}+\frac{1}{m_a m_b}
    \right) \right] \right.\nonumber\\
&+\mbox{}\gamma_a^0& m_a m_b  \left[-4 t+\frac{2 (m_a - m_b)}{ m_a^2
      m_b} t p^2 \right.\nonumber\\
&&+\left.\mbox{} \frac{t^3}{32 m_a^4}+\frac{t^2 p^2}{4
      m_a^4}+\frac{-m_a^3-m_a^2 m_b+m_a m_b^2+2 m_b^3}{2 m_a^4 m_b^3} t p^4
  \right] \nonumber\\
&+\mbox{} \gamma_b^0& m_a m_b \left[-4 t+\frac{2 (m_b - m_a)}{ m_a
      m_b^2} t p^2 \right.\nonumber\\
&&+\left.\mbox{} \frac{t^3}{32 m_b^4}+\frac{t^2 p^2}{4
      m_b^4}+\frac{2 m_a^3+m_a^2 m_b-m_a m_b^2- m_b^3}{2 m_a^3 m_b^4} t p^4
  \right] \nonumber\\
&+\mbox{} \gamma_a^0 \gamma_b^0& m_a m_b \left[ 4 t + \left( \frac{2}{m_a^2}+
      \frac{2}{m_b^2}-\frac{2}{m_a m_b} \right) t p^2\right.\nonumber\\
&&-\mbox{}t^3 \left(
      \frac{1}{32 m_a^4}+ \frac{1}{32 m_b^4}+\frac{1}{8 m_a^2 m_b^2}
    \right)-t^2 p^2 \left( \frac{1}{4 m_a^4}+\frac{1}{4 m_b^4}+\frac{1}{m_a^2 m_b^2}
    \right) \nonumber\\
&&-\left.\mbox{} t p^4 \frac{2 m_a^4+m_a^3 m_b+m_a^2 m_b^2+m_a m_b^3+2
        m_b^4} {2 m_a^4 m_b^4} \right]\nonumber\\
&+\mbox{}{\mbox{\boldmath $\gamma$}}_a \!\cdot \! {\mbox{\boldmath $\gamma$}}_b&\left. \left[ 2 t p^2+ \frac{t^2}{2}-\frac{t^2
    p^2}{4 m_a m_b}-\frac{t p^4}{m_a m_b} \right] \right\} u_a u_b
\label{eq:mitc1} 
\end{eqnarray}
and for the ${\cal C}_2$ piece:
\begin{eqnarray}
i \Im \left[{\cal M}_I^{C_2}\right]_t&=&\frac{i g^4}{16 \sqrt{t} (4 p^2+t)^2
  (m_a+m_b)m_a m_b}\; \bar{u}_a'{} \bar{u}_b'{} \left\{ \right.\nonumber\\
&\gamma_a^0&m_a m_b  \left[\frac{m_a^2+m_a
      m_b+m_b^2}{16 m_a^3 m_b} t (4 p^2+t)^2 \right]\nonumber\\
&+\mbox{}\gamma_b^0&m_a m_b 
  \left[\frac{m_a^2+m_a m_b+m_b^2}{16 m_a m_b^3} t (4 p^2+t)^2 \right]\nonumber\\
&-\mbox{}\gamma_a^0 \gamma_b^0 &\left[ \frac{m_a^4+m_a^3 m_b+2 m_a^2 m_b^2+m_a
    m_b^3+m_b^4}{16 m_a^2 m_b^2} t (4 p^2+t)^2 \right]\nonumber\\
&-\mbox{} {\mbox{\boldmath $\gamma$}}_a\! \cdot\! {\mbox{\boldmath $\gamma$}}_b
&\left. \left[\frac{m_a^2+m_a
      m_b+m_b^2}{16 m_a m_b} t (4 p^2+t)^2 \right] \right\} u_a u_b.
\label{eq:mitc2} 
\end{eqnarray}

In both equations above we kept only up to the powers of $t$ and $p^2$ that will be needed,
either in this section or in the next section, to obtain the leading two-particle exchange
potential. In this respect equation~(\ref{eq:mitc2}) does not contribute to the leading potential
just under scrutiny. To obtain the spectral density $\rho^{(4)}$, we must finally perform
the subtraction
\begin{equation}
\rho^{(4)} (s,t)=\frac{\Im \left[{\cal M}^{(4)}\right]_t-\Im \left[{\cal
      M}_I\right]_t}{2}.
\label{eq:roo}
\end{equation}

Recall that what enters $\rho^{(4)}$ is the real part of the $a_i$
in~(\ref{eq:disa}). Although the integrals that go in the $a_i$ are exactly
given in the appendix, the required subtraction~(\ref{eq:roo}) and final
Laplace transformation~(\ref{eq:pott}) demand that we here also expand
the integrands in a power series in $t$ and $p^2$. After some lengthy algebra, we arrive
at the final form for the imaginary part of $[{\cal M}]_t$
\begin{equation}
\Im \left[{\cal M} \right]_t = \Im \left[{\cal M}\right]_t^{odd} +\Im  \left[{\cal M}\right]_t^{even}
\end{equation}
where
\begin{eqnarray}
\Im \, \left[{\cal M}^{(4)}\right]_t^{odd} & = & \frac{g^4 m_a m_b}{4 (m_a+m_b) \sqrt{t} (4 p^2+t)^2}
\; \bar{u}_a'{} \bar{u}_b'{} \left\{  \right. \nonumber\\
& & \left[ 2 t-\frac{t p^2}{m_a m_b}+\frac{-3 m_a^4+2 m_a^2 m_b^2-3
    m_b^4}{4m_a^4 m_b^4} t p^4 \right.\nonumber\\
& & + \mbox{}\frac{-3 m_a^4-m_a^3 m_b+m_a^2 m_b^2-m_a m_b^3-3
    m_b^4}{8 m_a^4 m_b^4} t^2 p^2\nonumber\\
& &+\mbox{}\left.\frac{-3 m_a^4-m_a^3 m_b+m_a^2 m_b^2-m_a
      m_b^3 -3 m_b^4}{64 m_a^4 m_b^4} t^3 \right]\nonumber\\
&+\mbox{} \gamma_a^0 & \left[-2 t+\frac{m_a - m_b}{m_a^2 m_b} t p^2+\frac{3 m_a+4 m_b}{4
    m_a m_b^4} t p^4 \right.\nonumber\\
&  & +\mbox{}\frac{3 m_a^4+5 m_a^3 m_b+m_a^2 m_b^2-m_a m_b^3-m_b^4}{8
    m_a^4 m_b^4} t^2 p^2 \nonumber\\
& &+\mbox{}\left.\frac{3 m_a^4+5m_a^3 m_b+m_a^2 m_b^2-m_a m_b^3-
    m_b^4}{64 m_a^4 m_b^4} t^3 \right]\nonumber \\ 
&+\mbox{} \gamma_b^0 & \left[-2 t+\frac{m_b - m_a}{m_a m_b^2} t p^2+\frac{4 m_a+3 m_b}{4
    m_a^4 m_b} t p^4 \right.\nonumber\\
& &+\mbox{}\frac{-m_a^4-m_a^3 m_b+m_a^2 m_b^2+5 m_a m_b^3+3 m_b^4}{8
    m_a^4 m_b^4} t^2 p^2\nonumber\\
&&+\mbox{} \left. \frac{-m_a^4-m_a^3 m_b+m_a^2 m_b^2+5 m_a m_b^3+3
    m_b^4}{64 m_a^4 m_b^4} t^3 \right]\nonumber\\
&+\mbox{} \gamma_a^0 \gamma_b^0 & \left[ 2 t+\frac{m_a^2-m_a m_b+m_b^2}{m_a^2 m_b^2} t
  p^2 +\frac{t p^4}{m_a^2 m_b^2} \right.\nonumber\\
&  & +\mbox{}\frac{m_a^4+m_a^3 m_b+m_a^2 m_b^2+m_a
    m_b^3+m_b^4}{8 m_a^4 m_b^4} t^2 p^2\nonumber\\
& &+\mbox{}\left.\frac{m_a^4+m_a^3 m_b+m_a^2 m_b^2+m_a
    m_b^3+m_b^4}{64 m_a^4 m_b^4} t^3 \right]\nonumber\\
&+\mbox{} {\mbox{\boldmath $\gamma$}}_a \cdot {\mbox{\boldmath $\gamma$}}_b & \left[\frac{p^2 t}{m_a m_b}+\frac{t^2}{4
    m_a m_b}+\frac{m_a^2+4 m_a m_b+m_b^2}{2 m_a^3 m_b^3} t p^4\right.\nonumber\\
&  &+\left.\mbox{}\frac{3 m_a^2+5
    m_a m_b+3 m_b^2}{8 m_a^3 m_b^3} t^2 p^2+\frac{2 m_a^2 +m_a m_b+2 m_b^2}{32
    m_a^3 m_b^3} t^3 \right]\nonumber\\
& \left. \phantom{\left[\frac{p^2}{m_a}\right]} \right\} & u_a u_b \label{eq:carroim}
\end{eqnarray}
contains the odd powers of $\sqrt t$ and
\begin{eqnarray}
\Im\,\left[{\cal M}^{(4)}\right]_t^{even} & = &\frac{g^4}{4 \pi}\; \bar{u}_a'{} \bar{u}_b'{} \left\{ -\frac{ (m_a +m_b)
    \gamma_b^0 -m_b}{6 m_a^2 m_b} -\frac{ (m_a +m_b)
    \gamma_a^0 -m_a}{6 m_a m_b^2}+\frac{\gamma_a^{\mu}
    \gamma^b_{\mu}}{4 m_a m_b} \right.\nonumber\\
& & -\left.\mbox{}\frac{m_a m_b+\left( (m_a+ m_b)
      \gamma_a^0-m_a \right) \left( (m_a +m_b)
      \gamma_b^0-m_b \right)}{12 m_a^2 m_b^2} \right\} u_a u_b \label{eq:mres}
\end{eqnarray}
contains the even powers of $\sqrt t$. We did this separation to emphasize that, after
the subtraction in equation~(\ref{eq:roo}), only the term~(\ref{eq:mres})
survives to leading non-vanishing order. Indeed, equation~(\ref{eq:carroim}) coincides exactly with equation~(\ref{eq:mitc1}) if
we neglect terms beyond $t^2,p^4$, or $tp^2$.

 The final step involves the Laplace transformation indicated by
 equation~(\ref{eq:pott}). Using
the general formula
\begin{equation}
\int_0^{\infty}{t^n\:e^{-\sqrt{t}\,r}\,dt}=\frac{2\,(2
    n+1)!}{r^{2n+2}}
\end{equation}
we get
\begin{eqnarray}
U^{(4)}(r;s)&=&\frac{g^4}{16 \pi r^3}\; \bar{u}_a'{} \bar{u}_b'{} \left\{ -\frac{ (m_a +m_b)
    \gamma_b^0 -m_b}{6 m_a^2 m_b} -\frac{ (m_a +m_b)
    \gamma_a^0 -m_a}{6 m_a m_b^2}+\frac{\gamma_a^{\mu}
    \gamma^b_{\mu}}{4 m_a m_b}\right.\nonumber\\
& &-\mbox{}\left.\frac{m_a m_b+\left( (m_a+ m_b)
      \gamma_a^0-m_a \right) \left( (m_a +m_b)
      \gamma_b^0-m_b \right)}{12 m_a^2 m_b^2} \right\} u_a u_b \label{eq:u4psrel}
\end{eqnarray}
which leads, in the non relativistic limit and concentrating only on the spin-independent
terms of~(\ref{eq:u4psrel}), to the long-range attractive potential:
\begin{equation}
V_{nr}^{(4)}=-\frac{g^4}{64 \pi^3 r^3 m_a m_b} \, {\bf 1}^a_2 \otimes {\bf 1}^b_2 \label{eq:v4psnr}
\end{equation}
where this operator is supposed to act between two-component Pauli spinors.

\section{The Derivative Coupling}
\label{sec:der}

In this section we consider the interaction Lagrangian
\begin{equation}
{\cal L}^{der}_{int}=\frac{g}{2 m} \:\bar{\Psi} (x) \gamma_{\mu}\gamma^5 \Psi
  (x) \:\partial^{\mu} \Phi (x),
\end{equation}
which is how Goldstone bosons couple to fermions.

This derivative coupling leads to the same one particle exchange Feynman
amplitude as before and therefore to the same lowest order
potential~(\ref{eq:u2ps}) . Hence the iteration amplitude will be also
identical. However, the two particle exchange amplitude (see, Fig. 2) is
different because the Compton amplitude that goes into~(\ref{eq:m4c}) is
different. Indeed, the Compton amplitude, corresponding to Fig. 3, is now,
\begin{equation}
{\cal M}^C \left( k,k';P \right) =  g^2\; \bar{u} (p') \left[
  \frac{\rlap / {k}}{2\: p \cdot k}+\frac{\rlap / {k}'}{2\: p \cdot k'}-\frac{1}{m}\right] u(p).
\end{equation}

This amplitude differs from~(\ref{eq:comps}) by an extra term proportional to
$m^{-1}$.

We introduce this amplitude in~(\ref{eq:m4c}) and replace the massless
propagators by Dirac deltas to obtain the discontinuity function:
\begin{eqnarray}
\left[{\cal M}^{(4)}\right]_t=&-&\frac{i g^4}{8 \pi^2}\;\int{\!d\Phi\;\bar{u}_a'{} \left[
  \frac{p_a\cdot(k'-k)\rlap / {k}}{2\: p_a \cdot
    k' \: p_a\cdot k} \right] u_a\;\bar{u}_b'{} \left[\frac{p_b\cdot (k'-k)
    \rlap / {k}}{2\: p_b \cdot k' \: p_b\cdot k}\right]
u_b}\nonumber\\
&-&\frac{i g^4}{8 \pi^2}\;\int{\!d\Phi\,\left\{ \bar{u}_a'{} u_a\: \bar{u}_b'{} u_b \;
  \frac{1}{m_a m_b}-\frac{1}{m_a}\; \bar{u}_a'{} u_a\: \bar{u}_b'{} \left[ \frac{p_b \cdot
      (k'- k)}{2 \: p_b \cdot k \: p_b \cdot k '} \rlap / {k} \right]
    u_b\right.}\nonumber\\
&&-\mbox{} \left. \frac{1}{m_b}\; \bar{u}_a'{} \left[ \frac{p_a \cdot (k ' - k )}{2\:  p_a
        \cdot k \: p_a  \cdot k '} \rlap / {k} \right] u_a \bar{u}_b'{} u_b
  \right\}.%\label{eq:delm4}
 \end{eqnarray}

The first piece is exactly what we had in the last section. We call
$[\Delta {\cal M}^{(4)}]_{t}$ the extra added piece that involves the
integrations 
\begin{equation}
-\frac{i g^4}{8 \pi^2}\; \int{\!d \Phi \; \bar{u}_a'{} u_a\: \bar{u}_b'{} u_b \; \frac{1}{m_a
    m_b}}=-\frac{i g^4}{16 \pi m_a m_b} \; \bar{u}_a'{} u_a \: \bar{u}_b'{}
    u_b ,
\label{eq:dels1}
\end{equation}
and
\begin{eqnarray}
& & \frac{i g^4}{8 \pi^2 m_a}\; \int{\! d \Phi \; \bar{u}_a'{} u_a \: \bar{u}_b'{} \left[
    \frac{p_b \cdot (k ' -k)}{2\: p_b \cdot k \: p_b \cdot k '}
    \rlap / {k} \right]} u_b \quad + \quad \left( a \leftrightarrow b
    \right)\nonumber\\ 
& &=\frac{i g^4}{8 \pi^2 m_a}\; \int{\! d \Phi \; \bar{u}_a'{} u_a \: \bar{u}_b'{} \rlap / {k} u_b \,
  \frac{i \sqrt{t} \xi_b m_b x_b}{\frac{t}{2} \left(\frac{t}{4}+(\xi_b m_b
      x_b)^2 \right)}}\quad + \quad \left( a \leftrightarrow b
    \right)\nonumber \\ 
&  &= - \frac{g^4}{8 \pi m_a \sqrt{t} \xi_b m_b} \; \bar{u}_a'{} u_a \: \bar{u}_b'{}
\gamma_{\mu} u_b \, \left\langle \frac{x_b}{d_b}\: k^{\mu} \right\rangle \quad
    + \quad \left( a \leftrightarrow b \right) \nonumber\\
&  &= \frac{i g^4}{16 \pi m_a m_b \xi_b^2} \left( 1-\tau_b \arctan
  \left(\frac{1}{\tau_b} \right) \right) \: \bar{u}_a'{} u_a \:\bar{u}_b'{}
    u_b \quad + \quad \left( a \leftrightarrow b \right).
\label{eq:dels2} 
\end{eqnarray}

The last line in equation~(\ref{eq:dels2}) is reached by demanding Lorentz
covariance to write
\begin{equation}
\left\langle \frac{x_b}{d_b}\: k^{\mu} \right\rangle=a_b P_a^{\mu} + b_b P_b^{\mu}+c_b
  Q^{\mu},
\end{equation}
by solving for the coefficients as explained in the appendix, and by using the
Dirac equation.

Putting things together, 
\begin{eqnarray}
\left[ \Delta {\cal M}^{(4)} \right]_t&=&\frac{i g^4}{16 \pi m_a m_b} \;
    \bar{u}_a'{} u_a \: \bar{u}_b'{} \left[\frac{1}{\xi_a^2} \left(1-\tau_a \arctan \left(\frac{1}{\tau_a} \right)
    \right)\right.\nonumber\\
&&+\mbox{}\left.\frac{1}{\xi_b^2} \left(1-\tau_b \arctan \left(\frac{1}{\tau_b}
      \right) \right)-1 \right] u_a u_b \label{eq:dm4}
\end{eqnarray}

To leading order in $t$ and $p^2$ we have
\begin{equation}
\left[\Delta {\cal M}^{(4)}\right]_t=\frac{i g^4}{16 \pi m_a m_b} \; \bar{u}_a'{} u_a \: \bar{u}_b'{} u_b.
%\label{eq:afm4}
\end{equation}

In the nonrelativistic limit this contributes the quantity
\begin{equation}
\Delta V^{(4)}_{nr}=\frac{1}{4 \pi^2 r} \;\int_0^{\infty}{\!\Delta \rho \;e^{-\sqrt{t} r}\:
  dt}=\frac{g^4}{64 \pi^3 r^3 m_a m_b}
\end{equation}
which exactly cancels contribution~(\ref{eq:v4psnr}), i.e.
\begin{equation}
V^{(4)}_{der;nr}=V^{(4)}_{Y;nr}+\Delta V^{(4)}_{nr}=0+{\cal O} \left(r^{-4}
\right).
\end{equation}

Hence, to find the form for the potential in the case under scrutiny, we
must go to the next order in our series expansions. What we
need now is to consistently take into account the previously neglected
terms in the spectral density
\begin{equation}
\rho_{der}^{(4)}\equiv\frac{\left[{\cal M}_{Y}^{(4)}+\Delta {\cal M}^{(4)}-{\cal
  M}_{I}\right]_t}{2i}.
\end{equation}

So we collect the relevant pieces
in~(\ref{eq:carroim}),~(\ref{eq:mitc1}), and~(\ref{eq:mitc2}) in addition to 
\begin{equation}
\left[\Delta {\cal M}\right]_t^{(4)}  \sim \ldots - \frac{i g^4 \sqrt{t}}{16 m_a m_b}\; \bar{u}_a'{} \bar{u}_b'{}
\left( \frac{1}{4 m_a}+\frac{1}{4 m_b} \right) u_a u_b 
%\label{eq:delr4}
\end{equation}
which is the next to leading term in the expansion
of~(\ref{eq:dm4}). The result is 
\begin{eqnarray}
\rho^{(4)}_{der}& \equiv & \frac{\left[ {\cal M}^{(4)}_{Y}+\Delta {\cal
      M}-{\cal M}_I\right]_t }{2 i}\nonumber\\
& = & \frac{g^4 m_a m_b \sqrt{t}}{8 (m_a+m_b)}\; \bar{u}_a'{} \bar{u}_b'{} \, \left\{\right.\nonumber \\
& &-\mbox{}\frac{3 m_a^4+5 m_a^3 m_b+7 m_a^2 m_b^2+5 m_a m_b^3+3 m_b^4}{64 m_a^4 m_b^4}\nonumber\\
& &+ \mbox{}\gamma_a^0 \frac{3m_a^4+5 m_a^3 m_b-2 m_a m_b^3-3 m_b^4}{64 m_a^4
  m_b^4}\nonumber\\
& &+ \mbox{}\gamma_b^0 \frac{-3m_a^4-2 m_a^3 m_b+5 m_a m_b^3+3 m_b^4}{64 m_a^4
  m_b^4}\nonumber\\
& &+ \mbox{}\gamma_a^0 \, \gamma_b^0 \, \frac{3m_a^4+2 m_a^3 m_b+7 m_a^2
  m_b^2+2 m_a m_b^3+3 m_b^4}{64 m_a^4 m_b^4}\nonumber\\
& & \left.+ \mbox{}\mbox{spin-dependent terms of the kind}\; {\mbox{\boldmath $\gamma$}}_a
  \cdot {\mbox{\boldmath $\gamma$}}_b \right\}\, u_a u_b. \label{eq:ror4}
\end{eqnarray}

This spectral density nonetheless gives a vanishing spin-independent
potential in the static approximation, i.e.
\begin{equation}
V^{(4)}_{der;nr}=0+{\cal O} \left( r^{-5} \right).
\end{equation}

The first non-vanishing contribution to the spin-independent potential
arises from the part in the spectral density which is linear in $t$.
Indeed, the explicit form of the spectral density reads:
\begin{eqnarray}
\rho^{(4)}  \equiv && \frac{\left[{\cal M}^{(4)}_Y+\Delta{\cal M}^{(4)}\right]_t}{2 i}=\frac{g^4}{8 \pi }\;
\frac{1}{240 m_a^4 m_b^4}  \;\bar{u}_a'{} \bar{u}_b'{}\,\left\{ \right.\nonumber\\
& & 4\left[ \left( -3 m_a^4-2 m_a^3 m_b+2 m_a^2 m_b^2-2 m_a m_b^3-3 m_b^4
  \right) p^2 \right.\nonumber\\ 
& & \left.+\mbox{} \left(6 m_a^4+3 m_a^3 m_b+4 m_a^2 m_b^2+3 m_a m_b^3+6 m_b^4
  \right) t \right]\nonumber \\
&+ \gamma_a^0 & 2 (m_a+m_b) \left[ \left( 6 m_a^3-2 m_a^2 m_b+3 m_a m_b^2-4
    m_b^3 \right) p^2 \right.\nonumber\\ 
& & \left.+\mbox{} \left(-12 m_a^3+6 m_a^2 m_b-4 m_a m_b^2+3 m_b^3 \right)
  t \right]\nonumber\\
&+ \gamma_b^0 & 2 (m_a+m_b) \left[ \left( -4 m_a^3+3 m_a^2 m_b-2 m_a m_b^2+6
    m_b^3 \right) p^2 \right.\nonumber\\ 
& & \left.+\mbox{} \left(3 m_a^3-4 m_a^2 m_b+6 m_a m_b^2-12 m_b^3 \right)
  t \right]\nonumber\\
&+ \gamma_a^0 \gamma_b^0 & \left[ \left( 8 m_a^4+2 m_a^3 m_b-12 m_a^2 m_b^2+2
    m_a m_b^3+8 m_b^4 \right) p^2 \right.\nonumber\\ 
& &\left. \left.+\mbox{} \left(-6 m_a^4+2 m_a^3 m_b-9 m_a^2 m_b^2+2
    m_a m_b^3-6 m_b^4 \right) t \right]  \right\}\, u_a u_b
\label{eq:ror5}
\end{eqnarray}
where no iterated second order amplitude contributes to this order,
and where we picked the term proportional to $t$ in the expansion
of~(\ref{eq:dels1})~and~(\ref{eq:dels2}).

If we use now 
\begin{equation}
\bar{u}_a'{}\: \gamma_a^0 \:u_a \approx \bar{u}_a'{} \left( {\bf 1} +{\cal O} (p^2) \right) u_a
\label{eq:g0}
\end{equation}
and pass to the static limit, we find
\begin{equation}
\rho^{(4)}_{nr} = \frac{t}{32 m_a^2 m_b^2}\; {\bf 1}_a \otimes {\bf 1}_b
\end{equation}
which, upon Laplace transformation, leads to
\begin{equation}
V^{(4)}_{der;nr}=\frac{3 g^4}{128 \pi^3 m_a^2 m_b^2}\;
\frac{1}{r^5}\label{eq:vr5} 
\end{equation}
for the desired spin-independent long range potential. Note that, as opposed to
the Yukawa type coupling potential~(\ref{eq:v4psnr}), the derivative
interaction leads to a repulsive potential.

\section{Conclusions}
\label{sec:conc}

 Very light particles can mediate forces extending over
distances on the order of their Compton wavelength. If this
range is macroscopic, unpolarized  bulk
matter will only experience the effect of spin-independent
interactions. It is a well known fact that the Yukawa potential
due to pseudoscalar exchange depends on spin and as a
consequence no coherent effects do arise on a macroscopic scale,
unless of course our sample is polarized~\cite{moody}. However, residual
(Van der Waals type) forces may arise between macroscopic
bodies in the case of pseudoscalar mediated interactions,
due to the exchange of two quanta "at the same time". The
double helicity flip involved eventually makes the resulting
effective potential spin-independent~\cite{grifols}. 

 In the preceding sections we have established, with the help of the formalism
 developed by Feinberg and Sucher, the large
distance behavior of such residual forces, i.e. those
associated to double pseudoscalar exchange. We have
considered two different basic couplings of those scalars to
matter fermions. On the one hand we took the ordinary Yukawa
coupling (e.g., this is the way the Higgs particle couples to fermions)
and on the other we considered the derivative coupling
(e.g., the  axion-fermion interaction). Both couplings reduce
to the same spin-flip interaction in the static
non-relativistic limit. And both interactions also produce
identical one-particle exchange (spin-dependent) potentials.
In spite of this fact, we have explicitly shown that the
spin-independent two-particle exchange potential is
substantially different in both cases. Indeed, for Yukawa
coupling we derive a $1/r^3$ attractive long distance behavior whereas
for the derivative coupling the potential, that is now repulsive, falls off as
$1/r^5$. Since, as emphasized, double exchange will lead to
residual macroscopic effects, these effects will be quite
different in both cases. So, we have found still another
instance where the interaction of pseudoscalars to fermions can
be discriminated. Other places are, soft pion emission in
proton-proton scattering or axion bremsstrahlung in a
supernova core~\cite{rafa,turner}.  

 Of course, the effects just reported are extremely small for
the light scalars presently contemplated in particle physics,
such as the axion, and thus their experimental detection is
beyond reach of present technology. However, there is much
activity and interest on the experimental front and
experiments are designed and performed that explore the
sub-centimeter and sub-millimeter regime with an ever increasing
sensitivity~\cite{joshua,bordag}. And, on the theoretical side, the completion
of the Particle Physics Paradigm may still require new superlight
scalar particles to exist.  
    
\section*{Acknowledgments}
Work partially supported by the CICYT Research Project AEN95--0882. F.F. acknowledges the CIRIT for financial support.

\appendix
\section*{}
\label{sec:app}

\subsection*{The coefficients $a_i$}

The coefficients in the tensor decomposition~(\ref{eq:lor}) read:
\begin{eqnarray}
a_1 & = & \frac{t \left(I_1-2 I_5-I_6+4 y I_4-y^2 I_1-y^2 I_6 \right)}{16
  m_a^2 \xi_a^2 (1-y^2 )^2}\nonumber\\
a_2 & = & \frac{t \left(I_1-I_5-2 I_6+4 y I_4-y^2 I_1-y^2 I_5 \right)}{16
  m_b^2 \xi_b^2 (1-y^2 )^2}\nonumber\\ 
a_3 & = & \frac{t \left(- I_4-y I_1+2 y I_5+2 y I_6-3 y^2 I_4+y^3 I_1
  \right)}{16 b (1-y^2)^2}\nonumber\\
a_4 & = & -\frac{t \left(I_1-I_5-I_6+2 y I_4-y^2 I_1 \right)}{4
  (1-y^2)}\label{eq:sola} 
\end{eqnarray}
with $y\equiv\hat{{\bf p}}\cdot \hat{{\bf p}}'$, in terms of various angular
integrals in the set
\begin{eqnarray}
I_0 & \equiv & \left\langle \frac{1}{d_a d_b}\right\rangle \;\;\;\;I_1 \equiv
    \left\langle \frac{x_a x_b}{d_a
    d_b} \right\rangle\;\;\;\;I_2 \equiv \left\langle\frac{x_a^2}{d_a
    d_b}\right\rangle \nonumber \\
I_3 & \equiv & \left\langle\ \frac{x_b^2}{d_a d_b}\right\rangle \;\;\;\;I_4
    \equiv \left\langle \frac{x_a^2 x_b^2}{d_a
    d_b}\right\rangle  \;\;\;\;I_5 \equiv \left\langle\  \frac{x_a^3 x_b}{d_a
    d_b}\right\rangle  \nonumber\\
 I_6 & \equiv & \left\langle\ \frac{x_a x_b^3}{d_a d_b}\right\rangle
    .\label{eq:defij} 
\end{eqnarray}

These results are obtained after repeated contraction of~(\ref{eq:lor}) with
the independent momenta in our problem and after solving the resulting
algebraic system of equations. 

The angular integrals $I_i$ are given next. The explicit calculation of $I_i$
for $i<5$ is given in~\cite{fsd88} $I_5$ can be found in~\cite{fsd92} and $I_6$ follows trivially from $I_5$
by interchanging $a$ and $b$. For the first two integrals from the
set~(\ref{eq:defij}) one gets
\begin{equation}
I_0 = \frac{F_{+}+\pi N_{+}^{-1}}{2 \tau_a \tau_b}\;\;\;\;\;\;I_1 =
\frac{F_{-}+\pi N_{+}^{-1}}{2}
\end{equation}
where we have defined,
\begin{eqnarray}
F_{\pm} & = & \pm N_{-}^{-1} \arctan \left( \frac{N_{-}}{D_{+}}
\right)-N_{+}^{-1} \arctan \left( \frac{N_{+}}{D_{-}}\right)\nonumber\\
N_{+} & = & - \frac{i p s^{1/2}}{b}\nonumber \\
N_{-} & = & - \frac{i\sqrt{p^2 s+b y t}}{b}\nonumber \\
D_{\pm}& = &y \pm \tau_a \tau_b 
\end{eqnarray}
and the rest is given in terms of $I_0$ and $I_1$ by
\begin{eqnarray}
I_2 & = & \frac{1}{\tau_b} \arctan \left( \frac{1}{\tau_b}\right)- \tau_a^2
I_0 \nonumber \\
I_3 & = & \frac{1}{\tau_a} \arctan \left( \frac{1}{\tau_a}\right)- \tau_b^2
I_0 \nonumber \\
I_4 & = & 1-\tau_a \arctan\left( \frac{1}{\tau_a}\right)-\tau_b \arctan\left(
\frac{1}{\tau_b}\right)+\tau^2_a \tau^2_b I_0 \nonumber \\
I_5 & = & y \left(1-\tau_b \arctan \left(\frac{1}{\tau_b}\right) \right)-
\tau_a^2 I_1 \nonumber \\
I_6 & = & y \left(1-\tau_a \arctan\left( \frac{1}{\tau_a}\right) \right)- \tau_b^2 I_1.
\end{eqnarray}

\subsection*{Angular integrals entering the iterated amplitude}

 We start with~(\ref{eq:dl}). Its discontinuity $[L]_t$ has been obtained
in~\cite{fsd88}:
\begin{equation}
\left[{\cal L}\right]_t=i \left[ \frac{\pi}{l \sqrt{t}} \frac{1}{\sqrt{ \left(l_+^2-l^2
        \right) \left(l^2-l_-^2 \right)}} \right] \Theta \left(l_+^2-l^2
  \right) \Theta \left(l^2-l_-^2 \right)\label{eq:disl}
\end{equation}
with 
\begin{equation}
l_{\pm}^2=\frac{a' \pm b'}{2} \;\;\;\;\;a'=2p^2+t,\;\;\;
b'=\sqrt{t (4 p^2+t)}. 
\end{equation}
For $p^2=l^2$, simplifies to
\begin{equation}
\left. \left[{\cal L}\right]_t \right|_{p^2=l^2}=\frac{i \pi}{p^2\: t}.\label{eq:dislon}
\end{equation}

We turn now to~(\ref{eq:dv}). It has been calculated in~\cite{fsd92}. It is
symmetric in
${\bf p}$ and
${\bf p'}$. Hence it can be cast in the form
\begin{equation}
{\mbox{\boldmath ${\cal V}$}} = v {\bf P} \;\;\;\;with\;\;\;\; {\bf P} \equiv {\bf p}+{\bf p}\,{}'
  \label{eq:rv} 
\end{equation}
where $v$ is given by
\begin{equation}
v=\frac{1}{P^2} \left( (p^2+l^2){\cal L}-\frac{1}{4 p l} \ln
  \left[\frac{(p+l)^2}{(p-l)^2} \right] \right).
\end{equation}

The associated discontinuity is then 
\begin{equation}
\left[{\mbox{\boldmath ${\cal V}$}} \right]_t=\left[v \right]_t{\bf P}=\frac{p^2+l^2}{4 p^2+t}
\left[{\cal L} \right]_t{\bf P}.
\label{eq:disv}
\end{equation}

Finally let us discuss~(\ref{eq:dup}). This integral admits the general
decomposition
\begin{equation}
\Upsilon_{i j}=a P_i P_j+b Q_i Q_j+c \delta_{ij}\label{eq:rup}
\end{equation}
in terms of the vectors ${\bf P}$, defined above, and ${\bf 
Q}\equiv{\bf p}-{\bf p'}$. Clearly, we can write
\begin{eqnarray}
P_i P_j \Upsilon^{i j} & = & a \left( P^2 \right)^2+c\: P^2\nonumber\\
Q_i Q_j \Upsilon^{i j} & = & b \left( Q^2 \right)^2+c\: Q^2\nonumber\\
\delta_{ij}\Upsilon^{i j} & = & a P^2+b  Q^2+3 c
\end{eqnarray}
(we used: ${\bf P}\cdot{\bf Q}=0$). The scalar integrals on the left
are, respectively
\begin{eqnarray}
P_i P_j \Upsilon^{i j} & = & \int{\frac{d \Omega}{4 \pi} \frac{ \left({\bf P}
      \cdot {\bf l} \right)^2}{\left({\bf q}\,{}'{}^2\right) \left(
      {\bf q}^2\right)}}\nonumber\\
& = & (p^2+l^2) {\cal L}-\frac{p^2+l^2}{2 p l} \ln
\left[\frac{(p+l)^2}{(p-l)^2} \right]\nonumber\\
&& +\mbox{}\frac{4 p^2+t}{4
  p^2}+\frac{p^2+l^2}{16 p^3 l} t \ln
\left[\frac{(p+l)^2}{(p-l)^2} \right] \nonumber\\
Q_i Q_j \Upsilon^{i j} & = & \frac{1}{4} \left( -2+\frac{2 p^2+t}{p^2}+\frac{p^2+l^2}{4 p^3 l} t \ln
\left[\frac{(p+l)^2}{(p-l)^2} \right] \right)\nonumber\\
\delta_{ij}\Upsilon^{i j} & = & l^2 {\cal L}.
\end{eqnarray}

Now the three equations can be solved for $a$, $b$, and $c$. Recall that
we only need the discontinuity of~(\ref{eq:dup}), i.e. the discontinuities of
$a$, $b$, and $c$. They are, 
\begin{eqnarray}
\left[ a \right]_t& = & \frac{1}{4 p^2+t} \left( 2 \frac{(p^2+l^2)^2}{4 p^2+t}-l^2
\right) \left[{\cal L} \right]_t\nonumber\\
\left[ b \right]_t & = & \frac{\left[ c\right]_t}{t}\; \left[{\cal L} \right]_t\nonumber\\
\left[ b \right]_t & = & \left(l^2-\frac{(p^2+l^2)^2}{4 p^2+t} \right)\left[{\cal L} \right]_t. \label{eq:disup} 
\end{eqnarray}

\subsection*{Integrals $I_{c1}$ and $I_{c2}$}

Here we display the explicit solutions of~(\ref{eq:ic1})
and~(\ref{eq:ic2}) for
$n=0,2,4,6$: 
\begin{eqnarray}
I_{c1} (0) & = & 0 \label{eq:ic10}\nonumber\\
I_{c1} (2) & = & -\pi\nonumber\\
I_{c1} (4) & = & -\frac{\pi}{2} (4 p^2+t)\nonumber\\
I_{c1} (6) & = & -\frac{\pi}{8} \left( (4 p^2+t)^2+2  (2 p^2+t)^2 \right)
\end{eqnarray}
and,
\begin{eqnarray}
I_{c2} (0) & = & \pi \nonumber\\
I_{c2} (2) & = & \frac{\pi}{2} (2 p^2+t)\nonumber\\
I_{c2} (4) & = & \frac{\pi}{8} (8 p^4+12 p^2 t+3 t^2)\nonumber\\
I_{c2} (6) & = & \frac{\pi}{16} (16 p^6+48 p^4 t+30 p^2 t^2+5 t^3).
\end{eqnarray}

\newpage
\begin{center}
{\large{\bf Figure Captions}}
\end{center}

{\bf FIG. 1}
Lowest order scattering amplitude. Single pseudoescalar exchange.

{\bf FIG. 2}
Diagrams contributing to the ${\cal O} (g^4)$ terms of the potential.

{\bf FIG. 3}
Compton scattering amplitude diagrams.

\newpage
\begin{figure}
\begin{center}
\epsfig{file=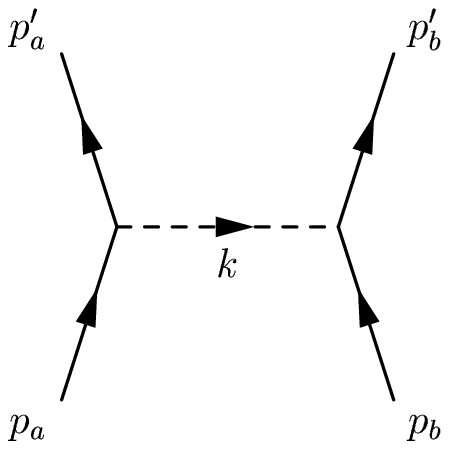,width=5.5cm,height=6cm}
\end{center}
\caption{}
\end{figure}

\begin{figure}
\begin{center}
\epsfig{file=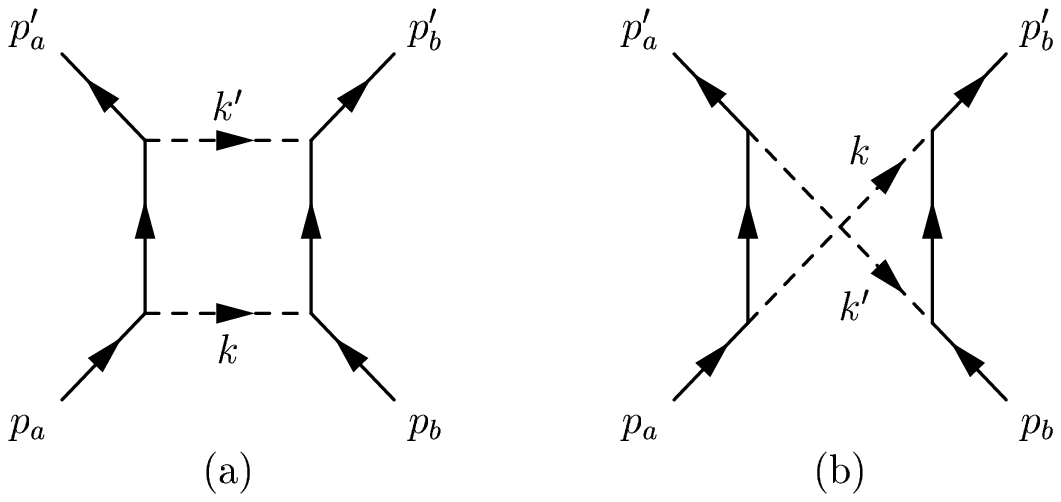,width=14cm,height=6.5cm}
\end{center}
\caption{}
\end{figure}

\begin{figure}
\begin{center}
\epsfig{file=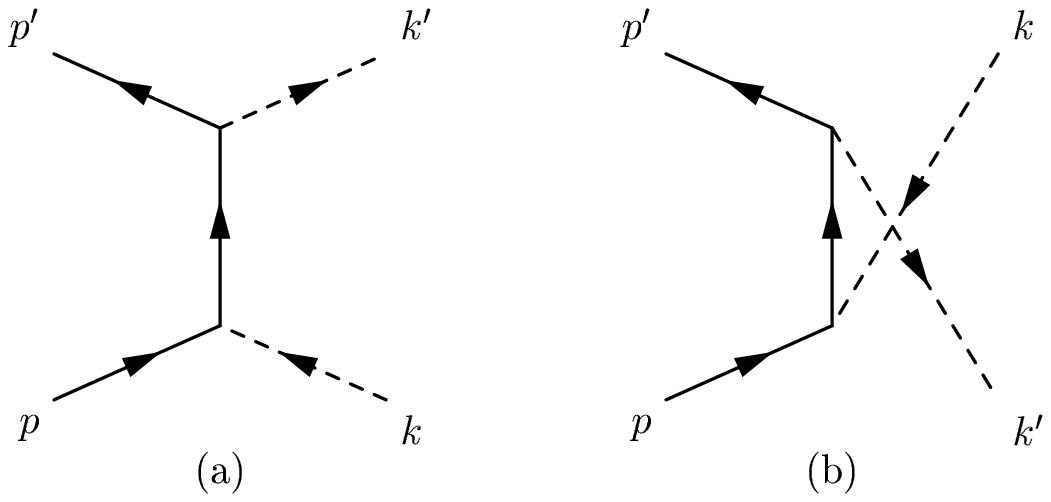,width=14cm,height=6.5cm}
\end{center}
\caption{}
\end{figure}

\end{document}